\documentclass[preprint]{aastex}

\usepackage{natbib}
\usepackage{graphicx}

\slugcomment{To appear in ApJLetters}

\shorttitle{Quasar superwinds}
\shortauthors{Barthel et al.}

\begin{document}

\title{Starburst-driven superwinds in quasar host galaxies}

\author{Peter Barthel and Pece Podigachoski}
\affil{Kapteyn Astronomical Institute, University of Groningen, The Netherlands.}
\email{pdb@astro.rug.nl}

\author{Belinda Wilkes}
\affil{Harvard-Smithsonian Center for Astrophysics, Cambridge, Massachusetts, USA.}

\and

\author{Martin Haas}
\affil{Astronomisches Institut, Ruhr Universit\"at, Bochum, Germany.}

\begin{abstract}

During five decades astronomers have been puzzled by the presence of
strong absorption features including metal lines, observed in the
optical and ultraviolet spectra of quasars, signalling in- and
outflowing gas winds with relative velocities up to several thousands of
km/sec.  In particular the location of these winds -- close to the
quasar, further out in its host galaxy, or in its direct environment --
and the possible impact on their surroundings have been issues of
intense discussion and uncertainty.  Using our Herschel Space
Observatory\footnote{Herschel is an ESA space observatory with science
instruments provided by European-led Principal Investigator consortia
and with important participation from NASA} data, we report a tendency
for this so-called associated metal absorption to occur along with
prodigious star formation in the quasar host galaxy, indicating that the
two phenomena are likely to be interrelated, that the gas winds likely
occur on the kiloparsec scale and would then have a strong impact on the
interstellar medium of the galaxy.  This correlation moreover would
imply that the unusually high cold dust luminosities in these quasars
are connected with ongoing star formation. Given that we find no
correlation with the AGN strength, the wind feedback which we establish
in these radio-loud objects is most likely associated with their host star
formation rather than with their black hole accretion.

\end{abstract}


\section{Introduction}

Soon after the discovery of quasi-stellar radio sources, QSRs, in 1963,
blue-shifted absorption features were found \citep{burbidge66} in the
optical (rest-frame ultraviolet) spectra of distant (high-redshift, $z >
1.5$) QSRs and of their radio-quiet counterparts, quasi-stellar objects
(QSOs).  The significantly lower redshift of most of these lines led to
the conclusion that they originate in gas along the line-of-sight,
opening an important new window to study the distribution and properties
of intervening galaxies and gas clouds \citep{blades88, weymann91}. 
Decades of research have revealed the diversity of these absorbers
\citep[e.g.,][]{bechtold02}, including intervening primordial gas
clouds, enriched haloes of intervening galaxies, and outflowing
circumnuclear gas winds in the quasar host galaxy, among others.  The
identity of one class of absorption lines, associated, metal-rich,
narrow absorption lines, remains elusive.  These associated metal
absorbers, seen mostly through ultraviolet C$^{3+}$ (\ion{C}{4}) and
Mg$^+$ (\ion{Mg}{2}) absorption, are narrow and strong, have velocity
widths less than $\sim300$ km/sec and rest-frame equivalent widths,
REWs, up to several~{\AA}, and occur \citep{foltz86, vester03} within a
velocity of a few thousand km/sec of the quasar.  The class is defined
by $\vert v \vert < 5000$ km/sec; both infall (negative $z_{\rm em}
-z_{\rm abs}$) and outflow (positive $z_{\rm em} -z_{\rm abs}$) occur. 
Detailed study \citep[e.g.,][]{williams75, sargent82, hamann01} of
certain associated metal absorber systems, using fine-structure line
information, has indicated that at least some are located at
$\sim$10~kpc of the central ionizing continuum source, however this is
not confirmed for the class as a whole. 

Quasar outflows are important as they remove gas from their host galaxy
and thereby the fuel for the AGN accretion as well as the material for
star formation.  Estimates of the mass of the associated outflowing gas
and its impact on the AGN host galaxy are a function of its distance and
coverage factor, both of which are unknown.  Nearby Mk\,231 represents
an intriguing case of this, so-called negative feedback.  Its luminous
far-infrared emission classifies this low-luminosity QSO as an
Ultraluminous Infrared Galaxy, ULIRG, in which the far-IR emission
implies a starburst galaxy nature \citep[e.g.,][]{surace98}.  Integral
field spectroscopy \citep{rupke13} has indicated the presence of massive
associated absorbing winds, covering wide angles on the kpc-scale,
suggesting sufficient outflowing material to have a significant impact
on the galaxy.  The ultimate driving mechanism for these winds could be
the starburst, the AGN, or their combination.  Mk\,231 may represent the
low luminosity equivalent of the Sloan Digital Sky Survey quasars for
which a statistical effect was reported \citep{shen12}, attributing
associated \ion{Mg}{2} absorption to host galaxy star formation, the
latter manifested by enhanced optical [\ion{O}{2}] emission.  In this
Letter we report further evidence favoring that associated metal
absorbers in quasars originate in starburst-driven winds.

\section{Observations and results}

Our group \citep{podi15, podi16a} has carried out far-infrared
imaging/photometry of the complete sample of 62 $z>1$ radio galaxies and
QSRs in the 3C catalogue, using guaranteed time on the Herschel Space
Observatory \citep{pilbratt10}, under projectcode GT1\_pbarthel\_1,
project title {\it The Herschel Legacy of distant radio-loud AGN}.  Our
spectral energy distribution (SED) modeling yields substantial cold
dust luminosities for a significant subset of QSRs and radio galaxies,
which we attribute to prodigious host galaxy star formation.  Standard
conversion formulae \citep{kennicutt98} yield star formation rates,
SFRs, in the range of $\sim 150 - 900$ M$_{\odot}$/year for one-third of
the sample, and upper limits of $\sim 200$ M$_{\odot}$/year for the
other two-thirds of the sample: in fact, several of the 3C objects
classify as (radio-loud) ULIRGs. 

Considering those nine sample 3C QSRs for which high resolution optical
(restframe ultraviolet) spectra are available in the literature, it is
striking that the two strongest star-formers, 3C\,298 (SFR=930
M$_{\odot}$/year) and 3C\,205 (SFR=700 M$_{\odot}$/year), are known
\citep{anderson87} to possess strong associated \ion{C}{4} absorption. 
In addition, another high SFR sample source, QSR 3C\,190 (SFR=470
M$_{\odot}$/year), had been reported \citep{stockton01} to possess
strong associated metal absorption (\ion{Mg}{2} -- unfortunately no
\ion{C}{4} data available) while having optical starburst signatures. 
These results motivated us to analyse our Herschel photometry of a
representative set of twelve $2 \lesssim z < 3$ 4C quasars which we had
added to the complete 3C sample with the aim of extending the redshift
coverage of powerful radio-loud AGN to $z \sim 3$.  Being more distant,
their radio luminosities are comparable to those of the 3C objects, and
their radio spectra and morphologies are also consistent with those of
the 3C sample.  The 4C-subsample Herschel photometry is published
elsewhere \citep{podi16b}.  The SED analysis led to an even more
striking result: only three of these sources are detected in the long
wavelength Herschel bands, hence experience extreme star formation, and
all three, 4C\,09.17 ($z$=2.111, SFR=1330 M$_{\odot}$/year), 4C\,24.61
($z$=2.330, SFR=1960 M$_{\odot}$/year), and 4C\,04.81 ($z$=2.586,
SFR=1570 M$_{\odot}$/year) have \citep{pdb90} strong associated
\ion{C}{4} absorption.  This suggests that the absorption is physically
connected to the star formation.  Dust component fitting of the near-IR
-- far-IR SEDs followed the techniques described in detail in
\citet{podi15}: \citet{hoenig10} warm dust tori reradiating the central
AGN illumination were combined with hot (1300K) black bodies, and with
cold grey bodies measuring the star formation luminosities.  Figure~1,
which can be found in the online material, presents the twelve SEDs and
dust component fits.  These fits permit extraction of the AGN as well as
star formation strengths. 

FIG~1: TWELVE 4C MIR-FIR SEDs

Adding the relevant 3C data from \citet{podi15}, Table~1 lists
SED-inferred AGN strengths, SFRs and upper limits for all high-redshift
3C and 4C QSRs for which high quality optical spectra of the \ion{C}{4}
region are available, with their associated \ion{C}{4} absorption
strengths and velocity information.  Figure~2 presents the SFRs as
function of the absorption line REWs. 

\newpage

TABLE~1: 3C/4C QSR, with SFR, AGN, CIV REW, and relative velocity Delta-v 

FIG~2: PLOT SFR vs. CIV REW

From Table-1 and Figure-2 it is seen that low-SFR 3C and 4C QSRs display
no, weak, intermediate and strong associated \ion{C}{4} absorption.  On
the other hand, the most prodigiously star-forming QSRs uniformly
display very strong absorption.  We stress that there is not a
one-to-one correspondence between SFR and absorption strength, as is for
instance illustrated by 3C\,191, the prototypical \citep{williams75,
hamann01} associated metal absorbing QSR with SFR upper limit
\citep{podi15} of 300 M$_{\odot}$/year.  However, considering the SFR
intervals $<400$, 400--1000, and $>1000$ M$_{\odot}$/year (indicated
with grey scales), the median \ion{C}{4} REW strengths are 0.9\AA,
1.9\AA, and 3.8\AA, with standard deviations 1.8\AA, 1.8\AA, and 2.6\AA,
respectively.  To further test the significance of the apparent trend,
we applied a median test.  Grouping the sources into high ($>$700
M$_{\odot}$/year) and low SFR, the probability of finding no high SFR
sources with REW$<$1.6\AA~ (the sample median) is 0.023.  While this is
a marginal result which should be confirmed using a larger sample, it is
consistent with the 3C\,190 result \citep{stockton01} and with the
statistical SDSS trend \citep{shen12}, thereby providing further
evidence for a relation between associated metal absorbers and star
formation. 

The sightline towards the compact optical/ultraviolet continuum source
is extremely narrow, so the chance of detecting absorption depends on
the coverage of the continuum source by absorbing material and should be
studied in a statistical sense.  The data indicate that the coverage
increases with increasing star formation rate.  This means that more of
the central continuum source is covered by absorbing material in the
strongest star-formers, such as 4C\,04.81, 4C\,24.61, and 4C\,09.17.  At
the same time, even in QSRs with a modest or small SFR, there is still a
chance of intercepting absorbing gas clouds, while unobstructed sight
lines (REW=0) also exist.  We conclude that the associated metal
absorption is likely to be directly related to the level of star
formation in these QSRs. 

\section{Discussion}

Recalling that the SFR=10 M$_{\odot}$/year starburst in the nearby
galaxy M\,82 occurs \citep{fenech08} within a 0.5~kpc region, that its
starburst-driven superwinds \citep{heckman90} extend over at least 1~kpc
from its center, and recalling the similar figures \citep{rupke13} in
starburst-QSO Mk\,231, the starbursts producing hundreds to thousands of
solar masses per year in the host galaxies of the 3C/4C QSRs likely
occur on scales of 1 -- 10 kpc.  Such kiloparsec-scale distances from
the central continuum source are consistent with the analyses
\citep{williams75, hamann01, sargent82} of some systems displaying
fine-structure lines, as mentioned earlier.  In fact, \ion{C}{4}
absorbing gas around starburst galaxies can extend up to distances as
large as 200~kpc \citep{bortha13}. 

The SFR-absorption trend for the radio-loud QSRs reported here suggests
therefore that starburst driven superwinds on the 1 -- 10 kpc scale are
responsible for the associated metal absorption.  Within that scenario,
the (generally adopted) assumption that cold dust emission is associated
with star formation is consistent.  The winds potentially constitute
important contributors to the chemical enrichment \citep{hamann99} of
the host galaxies and their environments.  It is interesting to note
that optical quasars displaying associated metal absorption tend to be
reddened with respect to unabsorbed quasars, consistent with their
inferred dusty hosts.  Following up on the \ion{C}{4} study of
\citet{vester03}, the first to report systematic reddening,
\citet{vdberk08} and \citet{shen12} measure $E(B-V) \sim 0.03$ in
quasars with associated \ion{Mg}{2} absorption.  \citet{vdberk08}
moreover found that the ionization of the associated absorbing gas cloud
is dependent on the relative velocity $\beta = \Delta{v}/c$ of the
cloud, in the sense that higher $\beta$ systems have lower ionization. 

Quantitative assessment (distance, mass, energetics) of the absorbing
clouds is difficult, as detailed information on their properties
(ionization, density, velocity structure, abundance) is lacking. 
Adopting steradian-scale coverage of the continuum source at distances
$\gtrsim$ 1~kpc, the absorbing gas must be in thin sheets, for all
reasonable values of neutral hydrogen columns N$_{\rm H}$.  As
\citet{hamann01} have shown for the case of 3C\,191, using high
resolution, high S/N Keck data, the absorber can have a multicomponent
nature.  These authors consider a blow-out leftover from a nuclear
starburst as a possible origin for the associated metal absorber in
3C\,191. 

In the meantime, we have other ways to address the wind properties.  For
the highest SFR objects in our sample, those having SFR $\sim$ 1000
M$_{\odot}$/year, Salpeter-type IMF starbursts lasting at least $10^7$
yrs (being the lifetime of supernova producing stars) will yield
\citep{woosley86} $\sim 0.1\%$ which is $\sim 10^7$ core-collapse
supernovae, with associated mass loss of $\sim 10^8$ M$_{\odot}$ and
total energy injection of $\sim 10^{58}$ ergs (adopting wind velocities
of a couple of thousand km/sec).  Alternatively, applying the superwind
model formulae \citep{heckman90} predicting the outflow properties from
the starburst's infrared luminosity while adopting a burst duration of
$10^7$ yrs, yields very similar numbers: instantaneous energy loss of
$\sim 10^{44}$ erg/sec, and integrated mass and energy loss $\sim 10^8$
M$_{\odot}$ and $\sim 10^{58}$ ergs, respectively.  These numbers are
broadly in line with the results obtained by \citet{hamann01} for
3C\,191.  We conclude that the feedback impact (per starburst; outflow
as well as inflow) is substantial.  As stated already, the detailed
properties of the absorption systems and their impact require further
high dispersion spectroscopic studies.  

Supporting evidence for the star formation origin of the kpc-scale
enriched gas in the host galaxies of these QSRs is provided by the
infrared spectroscopy of \citet{wilman00}.  These authors discovered
extended metal emission, in the form of [\ion{O}{3}], in the dusty 
Ly-$\alpha$ nebulae of several of the objects of the present study.
It is furthermore interesting to note that sample QSR 4C\,24.61, with
its extreme star formation rate, is the QSR displaying the highest
\citep{kronberg82} residual rotation measure, RRM, at radio wavelengths. 
This supports the suggestion \citep{watson91} that associated metal
absorbers are connected to the starburst-ionized ISM which is
responsible for the Faraday rotation of the radio emission. 

An issue of great importance concerns the nature of AGN feedback
\citep[e.g.,][]{fabian12}.  Given that the radio luminosities of the
QSRs under consideration, as well as their ultraviolet luminosities, are
comparable while their wind properties differ, we suspect that the wind
strengths are not driven by the AGN.  However, dust extinction is at
play and radio luminosity reflects AGN strength convolved with radio
source environment \citep[e.g.,][]{ba96}.  Integrated torus luminosities
therefore represent a more reliable measure of the AGN (accretion)
strength.  Taking data from Table~1, we show in Fig.~3 the 3C/4C AGN
strength as a function of the CIV equivalent widths: no correlation is
seen.  This strengthens our belief that -- at least for these radio-loud
objects -- the wind feedback is not governed by the AGN, but by the
contemporaneous star formation. 

FIG~3: PLOT AGN vs. CIV strength

Finally, we are likely witnessing the extreme forms of processes which
also occur in the low redshift universe.  Given the starburst nature
\citep{canalizo00, westhues16} of the $z=0.367$ FIR-ultraluminous QSR
3C\,48 and the $z=0.372$ FIR-luminous QSR 3C\,351, respectively, their
associated \ion{C}{4} absorption \citep{gupta05, mathur94} is consistent
with the same scenario.  The earlier suggestions \citep{heckman90}
concerning the starburst driven superwind nature of metal absorption
lines in FIR-bright quasars are consistent with our Herschel
observations. 

\section{Conclusions}

The occurrence of associated metal absorption, i.e., massive in- and
outflowing winds, in radio-loud quasars is seen to increase with
increasing star formation rate in the quasar host galaxies.  This
supports the view that these phenomena are physically related, is
consistent with the picture that long-wavelength FIR emission is
connected to star formation, and suggests that the wind feedback is
driven by the star formation and not by the AGN. 

\acknowledgments

PB acknowledges useful discussions with Fred Hamann, Michael Strauss,
and Sylvain Veilleux. A constructive referee report with valuable
suggestions is also gratefully acknowledged. The Herschel spacecraft
was designed, built, tested, and launched under a contract to ESA
managed by the Herschel/Planck Project team by an industrial consortium
under the overall responsibility of the prime contractor Thales Alenia
Space (Cannes), and including Astrium (Friedrichshafen) responsible for
the payload module and for system testing at spacecraft level, Thales
Alenia Space (Turin) responsible for the service module, and Astrium
(Toulouse) responsible for the telescope, with in excess of a hundred
subcontractors.  HCSS/HSpot/HIPE is a joint development by the Herschel
Science Ground Segment Consortium, consisting of ESA, the NASA Herschel
Science Center, and the HIFI, PACS and SPIRE consortia.

\begin{deluxetable}{lllcccc}
\tabletypesize{\scriptsize}

\tablecaption{\scriptsize Associated C\,IV absorption line properties, AGN strength,
 and SFRs for 9 3C and 12 4C QSRs. The 3C AGN and SFR data were taken from \citet{podi15};
 the 4C data result from the present analysis.
 The subsequent columns are: (1) 3C and 4C name; (2) IAU name (B1950); (3) redshift;
 (4) C\,IV rest-frame  equivalent width, in \AA~(from \citet{anderson87} and \citet{pdb90};
 in some cases these are multiple systems, and asterisks mark values which may be slightly
 higher as the emission line profile cannot be measured precisely);
 (5) absorption system velocity w.r.t. systemic, in km/sec (positive for infall);
 (6) warm dust luminosity inferred AGN strength, in $10^{12}$ L$_{\odot}$/year; 
 (7) cold dust luminosity inferred star formation rate and 3$\sigma$ upper limits,
 in M$_{\odot}$/year} 

\tablewidth{0pt}
\tablehead{
 \colhead{Name} & \colhead{IAU name} & \colhead{redshift} & \colhead{C\,IV REW} &
 \colhead{$\Delta{v}$} & \colhead{AGN strength} & \colhead{SFR} \\
 \colhead{} & \colhead{} & \colhead{} & \colhead{(\AA)} & \colhead{(km/sec)} &
 \colhead{($10^{12}$ L$_{\odot}$/yr)} & \colhead{(M$_{\odot}$/yr)}
 }
\startdata
3C\,9     & Q0017+154 & 2.014 & 0    & 0        & 17.1 & $<310$  \\
3C\,181   & Q0725+147 & 1.388 & 0.9  & $\sim$0  &  5.2 & $<150$  \\
3C\,191   & Q0802+103 & 1.954 & 6.1  & $-$610   & 17.0 & $<300$  \\
3C\,205   & Q0835+580 & 1.533 & 3.2  & 590      & 27.9 & 700  \\
3C\,268.4 & Q1206+439 & 1.400 & 1.9* & $-$1500  & 16.2 & $<140$  \\
3C\,270.1 & Q1218+339 & 1.519 & 6.2* & $-$2260  & 12.7 & 390  \\
3C\,298   & Q1416+067 & 1.440 & 4.5  & 120      & 29.2 & 930  \\
3C\,432   & Q2120+168 & 1.805 & 0.3  & $-$540   &  9.3 & 420  \\
3C\,454   & Q2249+185 & 1.758 & 0.5  & 1420     & 10.5 & 620  \\
 &  &  &  &  &   \\
4C\,$-$02.04 & Q0038$-$019 & 1.672 & 0    & 0       & 21  & $<240$  \\
4C\,17.09    & Q0109+176   & 2.155 & 5.4  & 190     & 8.1 & $<490$  \\
4C\,$-$01.11 & Q0225$-$014 & 2.038 & 0    & 0       & 7.7 & $<250$  \\
4C\,09.17    & Q0445+097   & 2.111 & 6.7  & $\sim$0 & 26  &  1330  \\
4C\,05.34    & Q0805+046   & 2.876 & 0.6  & $\sim$0 & 24  & $<750$  \\
4C\,28.40    & Q1606+289   & 1.981 & 7.0  & $-$910  & 9.5 & $<370$  \\
4C\,29.50    & Q1702+298   & 1.927 & 0    & 0       & 4.1 & $<250$  \\
4C\,47.48    & Q1816+475   & 2.223 & 0    & 0       & 15  & $<150$  \\
4C\,05.81    & Q2150+053   & 1.978 & 0.8  & 1210    & 14  & $<270$  \\
4C\,05.84    & Q2222+051   & 2.323 & 1.6  &$-$990   & 31  & $<540$  \\
4C\,24.61    & Q2251+244   & 2.330 & 3.8* &  270    & 33  &  1960  \\
4C\,04.81    & Q2338+042   & 2.586 & 2.4  & 330     & 37  &  1570  \\
\enddata
\end{deluxetable}

\clearpage

\begin{figure}
 \epsscale{0.80} 
 \plotone{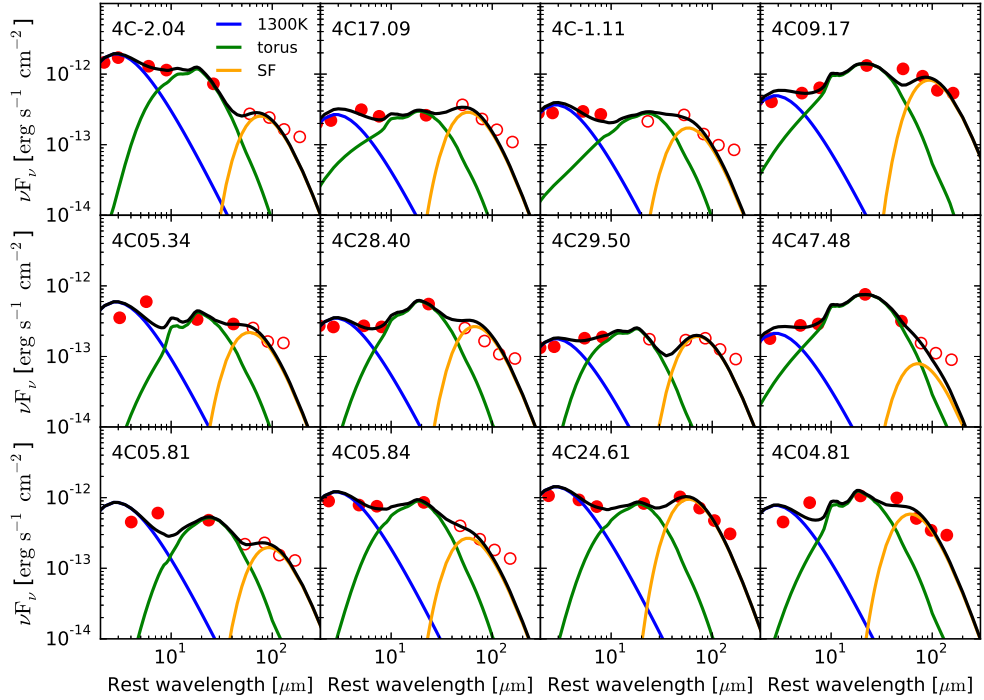}
 \caption{(Online material) NIR-MIR-FIR Spectral energy distributions for 
 twelve Herschel detected 4C QSRs, with dust component fits. Open circles
 represent 3$\sigma$ upper limits. The (green) warm torus models come from the 
 library of \citet{hoenig10}, and yield the AGN strengths. The (yellow) modified
 blackbodies fitted to the long  wavelength bands have fixed dust emissivity
 index $\beta=1.6$ (chosen on the basis of our 3C analysis); their luminosities
 were subsequently converted to SFRs (see the text). As discussed in detail
 in \citet{podi15}, QSRs generally show hot, 1300K dust (blue curves).
 The black lines represent the sum of the three components.}
\end{figure}

\begin{figure}
 \epsscale{0.80} 
 \plotone{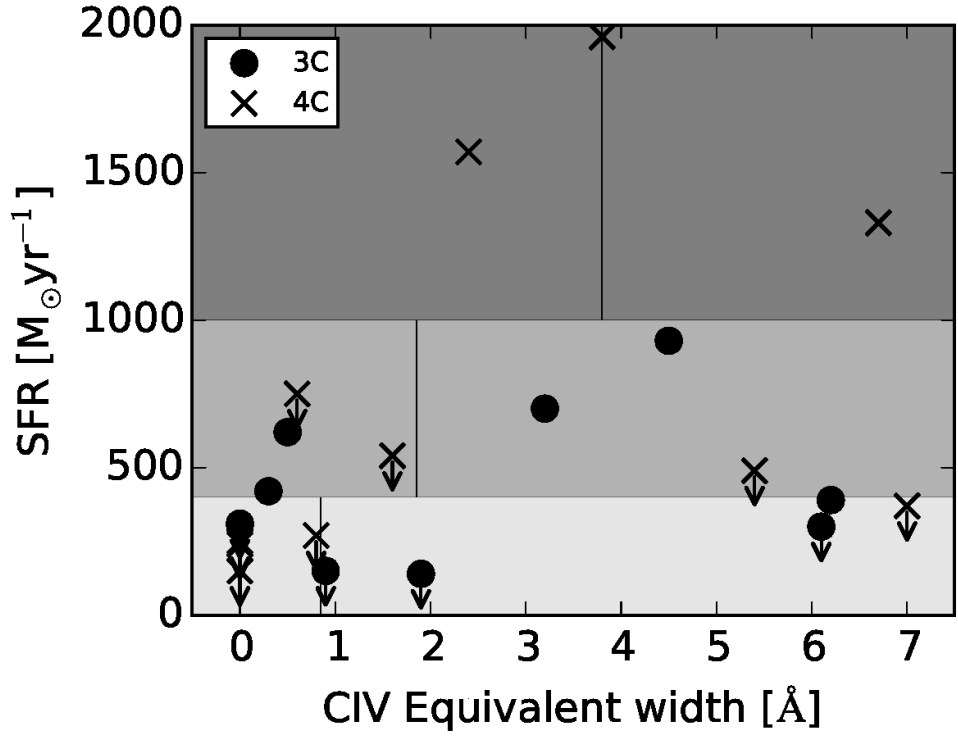}
 \caption{SFR vs. associated \ion{C}{4} absorption line strengths for 9 3C
 and 12 4C QSRs. The vertical lines indicate the median REW in the three separate
 SFR bins, marked with grey scales. These medians were obtained assuming all SFR
 upper limits to lie in the low SFR bin.}
\end{figure}

\begin{figure}
 \epsscale{0.80}
 \plotone{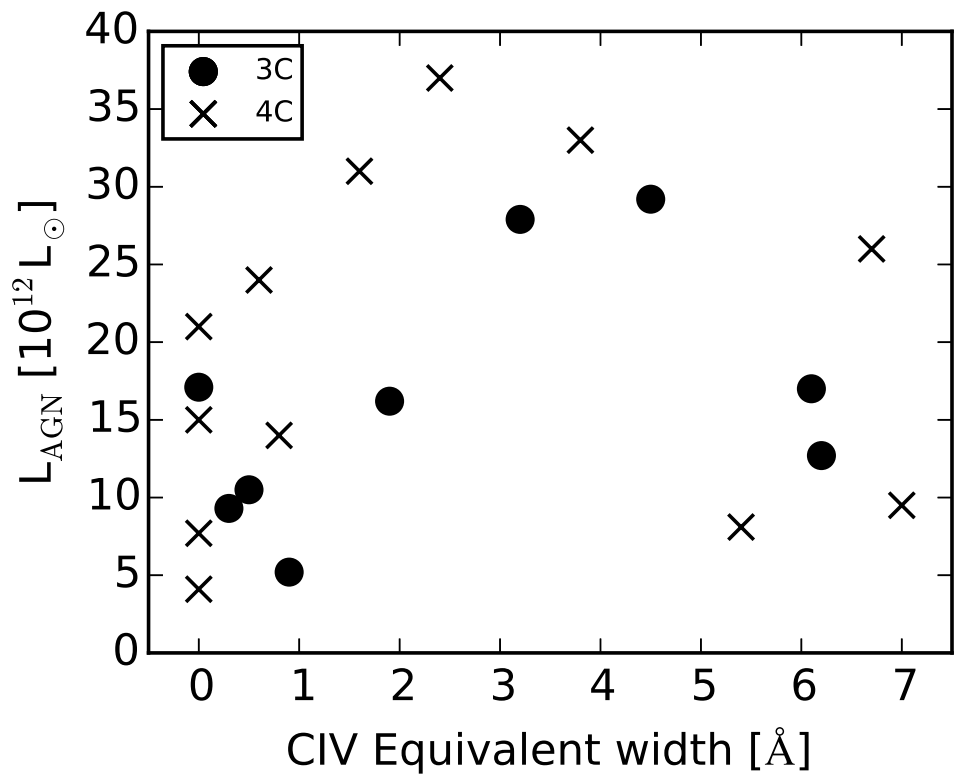}
 \caption{AGN strength vs. associated \ion{C}{4} absorption line strengths
 for 9 3C and 12 4C QSRs.}
\end{figure}


\begin{thebibliography}{}
\bibitem[Anderson et al.(1987)]{anderson87}
  Anderson, S.~F., Weymann, R.~J., Foltz, C.~B., Chaffee, Jr, F.~H. 1987,
 \aj, 94, 278
\bibitem[Barthel \& Arnaud (1996)]{ba96}
  Barthel, P.~D., \& Arnaud, K.~A. 1996, \mnras, 283, L45
\bibitem[Barthel et al. (1990)]{pdb90}                   
  Barthel, P.~D., Tytler, D.~R., Thomson, D. 1990, \aaps, 82, 339
\bibitem[Bechtold (2002)]{bechtold02}
  Bechtold, J. 2002, in "Galaxies at high redshift", eds.
  I.~P\'erez-Fournon, M.~Balcells, F.~Moreno-Insertis, and F.~S\'anchez
  (Cambridge University Press), p.~131
\bibitem[Blades et al.(1988)]{blades88}
  Blades, J.~C., Turnshek, D.~A., Norman, C.~A. (eds.), 1988, "QSO Absorption Lines:
  Probing the Universe", STScI Symposium Series Vol.2 (Cambridge University Press)
\bibitem[Borthakur et al.(2013)]{bortha13}
  Borthakur, S., Heckman, T., Strickland, D., Wild, V., Schiminovich, D. 2013, \apj, 768, 18
\bibitem[Burbidge et al.(1966)]{burbidge66}
  Burbidge, E.~M., Lynds, C.~R., Burbidge, G.~R. 1966, \apj, 144, 447
\bibitem[Canalizo \& Stockton (2000)]{canalizo00}
  Canalizo, G., \& Stockton, A. 2000, \apj, 528, 201
\bibitem[Fabian (2012)]{fabian12}              
  Fabian, A.~C. 2012, \araa, 50, 455
\bibitem[Fenech et al. (2008)]{fenech08}
  Fenech, D.~M., Muxlow, T.~W.~B., Pedlar, A., Beswick, R., Argo, M.~K. 2008,
  \mnras, 391, 1384
\bibitem[Gupta et al. (2005)]{gupta05}
  Gupta, N., Srianand, R., Saikia, D.~J. 2005, \mnras, 361, 451 
\bibitem[Foltz et al.(1986)]{foltz86}
  Foltz, C.~B., Weymann, R.~J., Peterson, B.~M., et al. 1986, \apj, 307, 504 
\bibitem[Hamann et al.(2001)]{hamann01}
  Hamann, F.~W., Barlow, T.~A., Chaffee, F.~C., Foltz, C.~B, Weymann, R.~J. 
  2001, \apj, 550, 142
\bibitem[Hamann \& Ferland (1999)]{hamann99}              
  Hamann, F., \& Ferland, G. 1999, \araa, 37, 487  
\bibitem[Heckman et al. (1990)]{heckman90}
  Heckman, T.~M., Armus, ~L., Miley, G.~K. 1990, \apj, 74, 833
\bibitem[H\"onig \& Kishimoto (2010)]{hoenig10}
  H\"onig, S.~F., \& Kishimoto, M. 2010, \aap, 523, A27
\bibitem[Kennicutt(1998)]{kennicutt98}
  Kennicutt Jr, R.~C. 1998, \araa, 36, 189
\bibitem[Kronberg \& Perry (1982)]{kronberg82}
  Kronberg, P.~P., \& Perry, J.~J. 1982, \apj, 263, 518
\bibitem[Mathur et al.(1994)]{mathur94}
  Mathur, S., Wilkes, B., Elvis, M., Fiore, F. 1994, \apj, 434, 493 
\bibitem[Pilbratt et al. (2010)]{pilbratt10}
  Pilbratt, G.~L., Riedinger, J.~R., Passvogel, T., et al. 2010, \aap, 518, L1
\bibitem[Podigachoski et al.(2015)]{podi15}
  Podigachoski, P., Barthel, P.~D., Haas, M., et al. 2015, \aap, 575, A80
\bibitem[Podigachoski (2016)]{podi16b}
  Podigachoski, P. 2016, "Star formation and AGN activity in
  distant massive galaxies", PhD Thesis, Univ. of Groningen 
\bibitem[Podigachoski et al. (2016)]{podi16a}
  Podigachoski, P., Rocca-Volmerange, B., Barthel, P., Drouart, G., Fioc, M.
  2016, \mnras, 462, 4183 
\bibitem[Rupke \& Veilleux (2013)]{rupke13}
  Rupke, D.~S.~N., \& Veilleux, S. 2013, \apj, 729, L27
\bibitem[Sargent et al.(1982)]{sargent82}
  Sargent, W.~L.~W., Young, P., Boksenberg, A. 1982, \apj, 252, 54 
\bibitem[Shen \& M\'enard (2012)]{shen12}         
  Shen, Y. \& M\'enard, B. 2012, \apj, 748, 131
\bibitem[Stockton \& Ridgway (2001)]{stockton01}
  Stockton, A., \& Ridgway, S.~E. 2001, \apj, 554, 1012
\bibitem[Surace et al.(1998)]{surace98}
  Surace, J.~A., Sanders, D.~B., Vacca, W.~D., Veilleux, S., Mazzarella, J.~M.
  1998, \apj, 492, 116
\bibitem[Vanden Berk et al. (2008)]{vdberk08}
  Vanden Berk, D., Khare, P., York, D.~G., et al. 2008, \apj, 679, 239
\bibitem[Vestergaard (2003)]{vester03}
  Vestergaard, M. 2003, \apj, 599, 116
\bibitem[Watson \& Perry (1991)]{watson91}
  Watson, A.~M., \& Perry, J.~J. 1991, \mnras, 248, 58   
\bibitem[Weymann et al. (1991)]{weymann91}
  Weymann, R.~J., Carswell, R.~F., Smith, M.~G. 1991, \araa, 19, 41
\bibitem[Westhues et al.(2016)]{westhues16}
  Westhues, C., Haas, M., Barthel, P., et al. 2016, \aj, 151, 120
\bibitem[Williams et al. (1975)]{williams75}
  Williams, R.~E., Strittmatter, P.~A., Carswell, R.~F., Craine, E.~R.
  1975, \apj, 202, 296
\bibitem[Wilman et al. (2000)]{wilman00}
  Wilman, R.~J., Johnstone, R.~M., Crawford, C.~S. 2000, \mnras, 317, 9 
\bibitem[Woosley \& Weaver (1986)]{woosley86}
  Woosley, S.~E., \& Weaver, T.~A. 1986, \araa, 24, 205
\end{thebibliography}
\end{document}